\documentclass[preprint,amsmath,amssymb]{revtex4}

\usepackage{graphicx}
\usepackage{subfigure}
\usepackage{dcolumn}
\usepackage{bm}

\begin{document}
\baselineskip=0.6cm


\title{
Quantum circuits for solving one-dimensional
Schr\"odinger equations
}

\author{K.~Nakao}
\author{A.~Matsuyama}

\affiliation{%
Department of Physics, Faculty of Science, Shizuoka University,
Shizuoka 422-8529, Japan
}%

\date{\today}

\begin{abstract}
We construct quantum circuits for solving one-dimensional
Schr\"odinger equations. Simulations of three typical examples,
i.e., harmonic oscillator, square-well and Coulomb potential,
show that reasonable results can be obtained with eight qubits.
Our simulations show that simple quantum circuits
can solve the standard quantum mechanical problems.
\end{abstract}

\maketitle

\section{Introduction}

Quantum computers have been one of the most growing fields
in computational physics last two decades.
Since Feynman suggested that
a quantum computer could possibly simulate quantum systems
more efficiently than a classical one \cite{fey},
a large amount of work has been
devoted to quantum algorithms and their experimental realizations.
This is because the quantum register (qubits) can store data
in the superposition of quantum states and operations of them
can be executed in parallel, which results in an exponential
reduction of the computation time and memories.
Among powerful applications of quantum algorithms are
Shor's factoring integers \cite{sho}
and Grover's searching databases \cite{gro}.

Since few quantum circuits are universal, i.e.,
any unitary operation on qubits can be constructed by those
universal quantum gates, it is, in principle,
possible to make appropriate quantum circuits
for calculating classical functions \cite{llo1,llo2}.
However, the efficiency of simulations with quantum circuits
is very much dependent on the dynamical system
under consideration.
Therefore we must find an efficient way of describing the system
and an efficient quantum simulation algorithm.

So far, there have been proposed several quantum algorithms
for simulating quantum mechanical systems.
Simulations of many body system have been reported, i.e.,
lattice-gas \cite{bog}, Heisenberg model \cite{rae, ovr1},
pairing model \cite{wu,ovr2} and Hubbard model
\cite{ovr2,abr1,som}.
Also quantum computations are expected to provide polynomial-time
simulation of chemical dynamics \cite{asp,kas}. 
These systems are suitable for the simulation with quantum algorithm,
since their quantum states are naturally represented by qubits,
i.e., $|0\rangle$/$|1\rangle$
of a qubit corresponds to the eigenstate of the number operator in the second
quantized formalism, or the up/down state of a spin, for example.
On the other hand, there have been few quantum simulations for particles
in real space, although a general algorithm was developed by Zalka \cite{zal}
and Wiesner \cite{wie}.
Recently, Benenti and Strini \cite{ben} simulated
time-evolution of a Gaussian wave packet
in the harmonic oscillator potential, and Oh \cite{oh}
calculated the ground state energy of
a displaced harmonic oscillator and a quartic anharmonic oscillator.

The purpose of this paper is to provide concrete examples of explicit simulation
of the one-dimensional Schr\"odinger equations of typical potentials, i.e.,
harmonic oscillator, square-well and Coulomb potential. We will explicitly
construct the quantum circuits for the calculation of the eigenvalues and
eigenstates of those Schr\"odinger equations.
In section 2, we will describe how to make quantum circuits for these three
examples. Several simulations will be reported in section 3, in which
outputs are compared with exact values. Section 4 is devoted to a summary.

\section{
Quantum algorithm for solving Schr\"odinger equations
}

In this section, we will briefly review the quantum algorithm
to solve Schr\"odinger equations \cite{zal,wie,nie}. 

\subsection{
Time-evolution of the quantum state
}

Let us consider the case where a single particle is moving
on a line under the potential $V(x)$.
The one-dimensional Schr\"odinger equation is
\begin{equation}
H|\psi\rangle
=\Bigl[ \frac{p^2}{2m} + V(x) \Bigr]|\psi\rangle
=i\hbar\frac{\partial}{\partial t}|\psi\rangle \ .
\label{eq1}
\end{equation}
Hereafter, we set the mass $m=1$ and Plank's constant
$\hbar=1$ for simplicity.
In the case of time-independent Hamiltonian,
the formal solution of Eq.(\ref{eq1}) is
\begin{equation}
|\psi(t)\rangle
= U(t)|\psi(0)\rangle
= e^{-iHt}|\psi(0)\rangle \ ,
\end{equation}
where $U(t)=e^{-iHt}$ is the unitary operator of time-evolution.
In order to calculate the time-evolution, firstly time-interval $t$
is divided into $n$ steps, i.e., $t=n\Delta t$, and then
each step is approximated by the second-order Trotter formula as
\begin{equation}
e^{-iH\Delta t}
= e^{-i(K+V)\Delta t}
= e^{-iV\Delta t/2}e^{-iK\Delta t}e^{-iV\Delta t/2}
+ O(\Delta t^3) \ ,
\end{equation}
where $K=p^2/2$ is the kinetic operator.
While it is straightforward to calculate $e^{-iV\Delta t/2}$
in the coordinate basis $|x\rangle$, it is preferable to
calculate $e^{-iK\Delta t}$ in the momentum basis $|p\rangle$.
The transformation of the basis is defined by
\begin{subequations}
\label{eq4}
\begin{eqnarray}
|p \rangle
&=& \int_{-\infty}^\infty dx |x \rangle\langle x | p \rangle 
= \int_{-\infty}^\infty d x ~e^{2 \pi i p x} |x \rangle \ ,
\\
|x \rangle
&=& \int_{-\infty}^\infty dp  |p \rangle\langle p | x \rangle
= \int_{-\infty}^\infty d p ~e^{-2 \pi i p x} |p \rangle \ ,
\end{eqnarray}
\end{subequations}
where we have used the convention for the later convenience.
The coordinate representation of the state vector $|\psi\rangle$ is
\begin{equation}
|\psi \rangle
= \int_{-\infty}^\infty  dx |x \rangle \langle x | \psi \rangle
= \int_{-\infty}^\infty dx \psi(x) |x \rangle \ ,
\quad
\psi(x)=\langle x|\psi\rangle \ ,
\end{equation}
and, the Fourier transformation of the wave function is
\begin{subequations}
\label{eq6}
\begin{eqnarray}
\psi(p) &=& \langle p|\psi\rangle
= \int_{-\infty}^\infty dx e^{-2 \pi i p x} \psi(x)
= U_{FT}^\dagger \psi(x) \ ,\\
\psi(x) &=& \langle x|\psi\rangle
= \int_{-\infty}^\infty dp e^{2 \pi i p x} \psi(p)
= U_{FT} \psi(p) \ .
\end{eqnarray}
\end{subequations}
Therefore, $x$-representation of the wave function
$|\psi(t)\rangle = U(t)|\psi\rangle$ is
\begin{eqnarray}
\lefteqn{
\langle x|e^{-iV\Delta t/2}e^{-iK\Delta t}e^{-iV\Delta t/2} |\psi \rangle
} \nonumber \\
& &
=e^{-iV(x)\Delta t/2}
\int_{-\infty}^\infty dp e^{2 \pi i p x}e^{-iK(p)\Delta t}
\int_{-\infty}^\infty dx' e^{-2 \pi i p x'}e^{-iV(x')\Delta t/2} \psi(x')
\nonumber \\
& &
=e^{-iV(x)\Delta t/2}
U_{FT} e^{-iK(p)\Delta t}
U_{FT}^\dagger e^{-iV(x')\Delta t/2} \psi(x') \ .
\end{eqnarray}

\subsection{
Discretization of the coordinate  and quantum Fourier transformation
}

We are interested in the bound state where the wave function
$\psi(x)$ is localized in some finite region.
The wave function $\psi(x)$ can be approximated on appropriate
mesh points $\{x_k\}$ in this region as
\begin{equation}
|\psi \rangle
= \sum_k \psi(x_k) |x_k \rangle \ .
\label{eq8}
\end{equation}
In order to carry out the Fourier transformation Eq.(\ref{eq6}),
we will employ quantum Fourier transformation (QFT).

The QFT is the unitary operation 
which transforms the basis
$\{|0\rangle,\ |1\rangle,\ldots\,|N\!\!-\!\!1\rangle \}$ to the new basis
$\{|\tilde{0}\rangle,\ |\tilde{1}\rangle,\ldots,
|\widetilde{N\!\!-\!\!1}\rangle \}$
such that
\begin{subequations}
\label{eq9}
\begin{eqnarray}
U_{QFT} :& &|j \rangle \to |\tilde{j}\rangle 
= \frac1{\sqrt{N}} \sum_{k =0}^{N-1} e^{2 \pi i j k / N} |k \rangle
\ ,  \\
U_{QFT}^\dagger : & &|\tilde{k} \rangle \to |k \rangle
= \frac1{\sqrt{N}} \sum_{j =0}^{N-1} e^{-2 \pi i j k / N}
|\tilde{j} \rangle \ .
\end{eqnarray}
\end{subequations}
By comparing with Eqs.(\ref{eq4}),
the bases $|k\rangle$ and $|\tilde{j}\rangle$
are identified with the coordinate basis $|x_k \rangle$
and the momentum basis $|p_j \rangle$
respectively.
Then inverse QFT of Eq.(\ref{eq8}) is
\begin{equation}
|\psi \rangle
= \sum_k \psi(x_k)|k \rangle
= \sum_{j,k} \psi(x_k) \frac1{\sqrt{N}}e^{-2 \pi i j k / N} |\tilde{j} \rangle 
= \sum_j \psi(p_j) |\tilde{j} \rangle \ .
\end{equation}
Therefore the inverse QFT changes the $x$-representation of $\psi(x)$
to the $p$-representation $\psi(p)$ \cite{zal,wie}.

By making suitably scaling and shifting the coordinate, 
we will choose the $x$-space interval $[-1/2,1/2]$ for simplicity,
and $N$ equally spaced
mesh points, i.e.,\ $x_k=k/N-1/2, \ (k=0,1,\ldots,N-1)$. 
Accordingly, the $p$-space mesh points are taken as
$p_j=2\pi(j-N/2), \ (j=0,1,\ldots,N-1)$.
In this case, QFT Eqs.(\ref{eq9}) are
\begin{subequations}
\label{eq11}
\begin{eqnarray}
|\tilde{j}\rangle
&=&
\frac1{\sqrt{N}} \sum_{k =0}^{N-1} e^{2 \pi i (j-N/2)(k/N-1/2)} |k \rangle \ ,  \\
|k \rangle
&=& \frac1{\sqrt{N}} \sum_{j =0}^{N-1} e^{-2 \pi i (j-N/2)(k/N-1/2)}
|\tilde{j} \rangle \ .
\end{eqnarray}
\end{subequations}
The phase factor becomes $e^{2\pi ijk/N}e^{-\pi i(k+j)}e^{2\pi i N/4}$.
The constant factor $e^{2 \pi i N/4}$ can be absorbed in the bases.
Redefining the new bases
\begin{equation}
|k'\rangle = e^{-\pi i k} |k\rangle =(-1)^k|k\rangle,\quad 
|\tilde{j}'\rangle = e^{\pi i j} |\tilde{j}\rangle =(-1)^j|\tilde{j}\rangle
\ ,
\end{equation}
the standard QFT Eqs.(\ref{eq9}) can be satisfied.
Therefore, in executing the practical calculation with the standard QFT,
the wave function should also be redefined as
\begin{equation}
|\psi \rangle
= \sum_k \psi(x_k)|k \rangle
= \sum_k \tilde{\psi}(x_k)|k' \rangle \ ,\ \tilde{\psi}(x_k)=(-1)^k\psi(x_k)
\ .
\end{equation}

The distribution of mesh points $\{x_k\}$ described above
is not exactly symmetric with respect to $x=0$.
This may cause some numerical inconvenience for symmetric potentials.
The mesh point $x_k=0$ is also not suitable for Coulomb potential.
Therefore it is convenient to prepare another distribution
which is exactly symmetric and does not contain the point $x=0$.
They are
\begin{equation}
x_k=\frac{k}{N}-\Bigl( \frac12-\frac1{2N} \Bigr), \quad 
p_j=2\pi \Bigl[j-\Bigl( \frac{N}{2}-\frac12 \Bigr)\Bigl] \ .
\label{eq14}
\end{equation}
In this case, the Eqs.(\ref{eq11}) become
\begin{subequations}
\begin{eqnarray}
|\tilde{j}\rangle 
&=&
\frac1{\sqrt{N}} \sum_{k =0}^{N-1} e^{2 \pi i (j-N/2+1/2)(k/N-1/2+1/2N)}
|k \rangle \ , \\
|k \rangle
&=& \frac1{\sqrt{N}} \sum_{j =0}^{N-1} e^{-2 \pi i (j-N/2+1/2)(k/N-1/2+1/2N)}
|\tilde{j} \rangle \ .
\end{eqnarray}
\end{subequations}
Then, with new bases
\begin{equation}
|k'\rangle = e^{-2\pi i(1/2-1/2N)k} |k\rangle ,\quad 
|\tilde{j}'\rangle = e^{2\pi i(1/2-1/2N)j} |\tilde{j}\rangle \ ,
\end{equation}
the standard QFT is satisfied. The wave function is accordingly redefined as
\begin{equation}
\tilde{\psi}(x_k)=e^{2\pi i(1/2-1/2N)k}\psi(x_k) \ .
\end{equation}
This distribution will be employed in the next section.

\subsection{
Phase estimation and eigenfunction
}

If we take the initial state $|\psi(0)\rangle$ to be
an eigenstate $|u_k\rangle$ of the Hamiltonian $H$
with eigenvalue $E_k$, i.e., 
$ H |u_k\rangle = E_k |u_k\rangle$,
then $U(t)|u_k\rangle = e^{-iE_k t}|u_k\rangle$ and the energy $E_k$
can be calculated by the phase estimation algorithm \cite{abr2}.
Since the phase estimation algorithm finds the eigenvalue
$e^{2\pi i \phi}$ of the unitary operator $U(t)$, the energy
eigenvalue $E_k$ is given by
\begin{equation}
E_k=-2\pi \phi/t, \quad (0 \le \phi < 1) \ .
\end{equation}
In order to make the energy eigenvalue negative, it is necessary
to shift the Hamiltonian by an appropriate constant value.
One should also choose the evolution time $t$
such that the searched energy range is $(E_{max}-E_{min})=2\pi/t$.
Since the phase estimation algorithm gives us the same $E_k$
periodically, we should be careful about the situation
where different energy states may contribute the same energy phase.

The phase estimation algorithm consists of two kinds of registers,
i.e., the first register is work qubits for storing the phase of
the unitary operator $U(t)$, and the second register is the simulation
qubits for representing the quantum state. The total state is
a tensor product of work qubits and simulation qubits.
In general case, the initial state $|\psi(0)\rangle$ is
written by the superposition of eigenstates of $H$ as
\begin{equation}
|\psi(0)\rangle = \sum_k c_k |u_k\rangle \ .
\label{eq19}
\end{equation}
Thus, the total state is also the superposition of the
tensor products, and one can find the coefficient $c_k$
by the projection operator of the corresponding work qubits
\cite{ovr1,ovr2}.
In order to execute efficient simulations, the initial state
should be prepared in such a way that it has an appreciable
overlap with the eigenstate $|u_k\rangle$
which we are searching.

\section{
Simulations of typical examples
}

In this section, we will show the practical way
of constructing quantum circuits for three
typical examples, i.e.,
harmonic oscillator potential, square-well potential,
and Coulomb potential ($S$-wave).
The range of the coordinate $x$ is fixed to $[-1/2, 1/2]$,
and we will choose the strength of the potential such that
the wave function is localized in this range.
We will employ $w$ work qubits (first register) and $s$
simulation qubits (second register). Thus the dimensions of
the work and simulation bases are $N_w=2^w$ and $N_s=2^s$
respectively. Total number of qubits is $q=w+s$
and the dimension is $N_q=2^q=N_w N_s$.
Then, we will prepare equally spaced $N_s$ mesh points for
$-1/2 \le x \le 1/2$,
and $N_w$ energy points for $2\pi/t=(E_{max}-E_{min})$
with energy step size $\Delta E=2\pi/t/N_w$.

In the practical calculations of the following examples,
we set $w=s=4$. The typical energy scale is $10^2$,
and simulations give good convergence with divided
time interval $\Delta t=t/n \simeq 10^{-3}$.
We have carried out several calculations with more qubits
and time-steps, and certainly obtained improved results,
although the qualitative features remain the same.
Therefore we will show the results of simulations with $w=s=4$,
which can be executed within reasonable computer resources.
In our experience of numerical calculations on the ordinary
(classical) computer, $2^4=16$ mesh points or bases are
sufficient to obtain ground and a few excited states
in one-dimensional potential.
So it is expected that simulations with $w=s=4$
could give us outputs with more or less similar accuracy.

The quantum circuits for the quantum Fourier transformation
(QFT) and the phase estimation are well known and
detailed descriptions are given in Ref.\cite{nie}
for example.
Thus we will not repeat the explanation of these circuits.
In Ref.\cite{nie}, one can also find how efficient is
the quantum simulation algorithm.

In the following subsections, we will explicitly construct
quantum circuits of the time-evolution operator
$U(t)=e^{-iH\Delta t}$, execute simulations 
with appropriate initial states, and compare the outputs
with exact values.
For these examples, quantum circuits can be constructed
only by single- and two-qubit operators. Furthermore,
ancillary qubits calculating the potential term are not
necessary. The phase-evolution due to the potential term
is implemented directly in the time-evolution circuit.

\subsection{Kinetic energy term}

Let us begin with the quantum circuit of
the common kinetic energy term $e^{-iK \Delta t}$.
The time-evolution operator of the kinetic term is
\begin{equation}
e^{-iK \Delta t} |p_j \rangle 
= e^{-i \frac12 p_j^2 \Delta t} |p_j \rangle 
= e^{i\alpha (\frac{j}{N_s}-\frac12)^2} |p_j \rangle
\ ,\ \alpha=-(2\pi N_s)^2 \Delta t/2 \ .
\end{equation}
The integer $j$ is represented by the binary form as,
\begin{equation}
j
= \sum_{n=1}^s j_n 2^{s-n}
=j_1 2^{s-1}+j_2 2^{s-2}+\ldots + j_s 2^0
=j_1j_2\ldots j_s({\rm binary}) \ .
\end{equation}
Thus $j/N_s \in [0,1]$ is the binary fraction
\begin{equation}
j/N_s 
= \sum_{n=1}^s j_n 2^{-n}
=j_1 2^{-1} + j_2 2^{-2} + \ldots +j_s 2^{-s}
=0.j_1j_2 \ldots j_s({\rm binary}) \ .
\end{equation}
Therefore
\begin{equation}
(j/N_s -1/2)^2=(\sum_n j_n 2^{-n} -1/2)^2
=(\sum_n j_n 2^{-n})^2 - \sum_n j_n 2^{-n} +1/4 \ .
\label{eq23}
\end{equation}
The computational basis $|p_j \rangle$ is the direct product of
$s$ qubits
\begin{equation}
|p_j \rangle = |j_1j_2\ldots j_s \rangle
=|j_1 \rangle \otimes |j_2 \rangle \ldots \otimes |j_s \rangle
\ .
\end{equation}
The last term of Eq.(\ref{eq23}) simply multiplies a constant factor
$e^{i\alpha/4}$ on one of the simulation qubits,
$|j_1 \rangle$ for example. The operation of the second term is
\begin{equation}
e^{-i\alpha \sum_n j_n 2^{-n}}|j_1j_2\ldots j_s \rangle
= \bigotimes_n e^{-i \alpha j_n 2^{-n}} |j_n \rangle
= \bigotimes_n R(- \alpha 2^{-n}) |j_n \rangle
\ ,
\end{equation}
where $R(\theta)$ is the single-qubit operator
rotating the phase of $|1\rangle$ by $\theta$,
i.e.,
\begin{equation}
R(\theta) = 
\left(
\begin{array}{cc}
1 & 0 \\
0 & e^{i\theta}
\end{array}
\right)
\ .
\end{equation}
The first term of Eq.(\ref{eq23}) is written by
\begin{equation}
e^{i\alpha \bigl(\sum_m j_m 2^{-m}\bigr)
\bigl(\sum_n j_n 2^{-n}\bigr)}
=\exp\Bigl(
i\alpha \sum_n j_n^2 2^{-2n}
+ 2i\alpha\sum_{m \ne n} j_m j_n 2^{-m-n}
\Bigr) \ .
\end{equation}
Since $j_n^2=j_n$,
$e^{i\alpha j_n 2^{-2n}}$ is a single-qubit
operator given by $R(\alpha 2^{-2n})$. On the other hand,
\begin{equation}
j_m j_n =
\left \{
\begin{array}{ll}
0 & \qquad j_m=0\ {\rm or} \ j_n=0 \\
1 & \qquad j_m = j_n = 1 \quad ,
\end{array} 
\right.
\end{equation}
$e^{2i\alpha j_m j_n 2^{-m-n} }$
is given by a two-qubit operator $A$ as
\begin{equation}
A =
\left(
\begin{array}{c c c c}
1 & 0 & 0 & 0 \\
0 & 1 & 0 & 0 \\
0 & 0 & 1 & 0 \\
0 & 0 & 0 & e^{2 i \alpha 2^{-m-n}} 
\end{array}
\right)
\ .
\end{equation}
This two-qubit operator acting on $|j_mj_n \rangle$
can be represented by the controlled-$U$
($CU$) operation shown in Fig.1.

\begin{figure}[h t b p]
\begin{center}
\unitlength 0.1in
\begin{picture}( 25.8500, 10.8500)( 5.0,-16.0)
%
\special{pn 8}%
\special{sh 0.600}%
\special{ar 2000 600 50 50  0.0000000 6.2831853}%
%
\special{pn 8}%
\special{pa 1800 1200}%
\special{pa 2200 1200}%
\special{pa 2200 1600}%
\special{pa 1800 1600}%
\special{pa 1800 1200}%
\special{fp}%
%
\special{pn 8}%
\special{pa 1200 600}%
\special{pa 1950 600}%
\special{fp}%
\special{pa 2050 600}%
\special{pa 2800 600}%
\special{fp}%
\special{pa 2800 1400}%
\special{pa 2200 1400}%
\special{fp}%
\special{pa 1800 1400}%
\special{pa 1200 1400}%
\special{fp}%
%
\special{pn 8}%
\special{pa 2000 650}%
\special{pa 2000 1200}%
\special{fp}%
\put(8.0000,-6.0000){\makebox(0,0){$|j_m\rangle$}}%
\put(8.0000,-14.0000){\makebox(0,0){$|j_n\rangle$}}%
\put(20.0000,-14.0000){\makebox(0,0){$U$}}%
\end{picture}%
\\

FIG\ 1:\ Quantum circuit $CU$
\end{center}
\end{figure}
\setcounter{figure}{1}

In the $CU$ circuit,
the single-qubit operator $U$ is applied to the target qubit
when the control qubit is set to $|1 \rangle$.
In this case, $U=R(2\alpha 2^{-m-n})$. For $s$ simulation qubits,
the number of two-qubit operator is ${}_sC_2=s(s+1)/2$,
and one should apply $CU$ gates successively.

\subsection{Harmonic oscillator potential}

The Hamiltonian of the harmonic oscillator is
\begin{equation}
H=\frac12 p^2+\frac{\omega^2}{2}x^2 \ .
\end{equation}
Since the potential term is the same quadratic form
as the kinetic term,
the quantum circuit of the time-evolution $e^{-iV\Delta t/2}$
is the same as the kinetic energy term,
except that the strength $\alpha=-(2\pi N_s)^2 \Delta t/2$
is replaced by $\beta=-\omega^2 \Delta t/4$.
Then, the calculation of the time-evolution operator is
\begin{equation}
e^{-iV\Delta t/2}e^{-iK\Delta t}e^{-iV\Delta t/2}
|\psi \rangle
=e^{-iV\Delta t/2} U_{QFT}
e^{-iK\Delta t} U_{QFT}^\dagger
e^{-iV\Delta t/2} |\psi \rangle
\ .
\end{equation}

We have chosen the potential strength parameter $\omega$ such that
both $x$- and $p$-space wave functions $\psi(x)$ and $\psi(p)$
are well localized in the chosen finite interval and transformed
accurately by the QFT. In practice, optimal parameter $\omega$
is given by
\begin{equation}
\omega/2=(2\pi N_s)^2/(2\omega), \quad \omega=2\pi N_s \ .
\end{equation}
For $s=4$, $\omega \simeq 100.53$ and we fixed
$\omega=100$ in the following calculations.

We will show the probability spectrum $|c_k|^2$
of Eq.(\ref{eq19}) as a function of the energy $E$ in Figs.2
for the initial states $\psi_0(x)=e^{-\omega x^2/2}$
and $\psi_0(x)=xe^{-\omega x^2/2}$
(the normalization factor will be omitted hereafter).
Since these initial states are exact eigenstates,
the outputs are good check for the simulation.
The energy spectrum of Fig.2(a) clearly shows the sharp peak around
the exact value $E_0=\omega/2=50$. The numerical value is
$|c|^2=0.915$ at $E=52.4$.
Fig.2(b) shows the result of the first excited state and
output of simulation is $|c|^2=0.699$ at $E=157$.
These examples show that our simulations work fairly well.

\begin{figure*}[h t b p]
\begin{center}
\begin{tabular}[t]{cc}
\subfigure[]{\includegraphics[width=.4\textwidth]{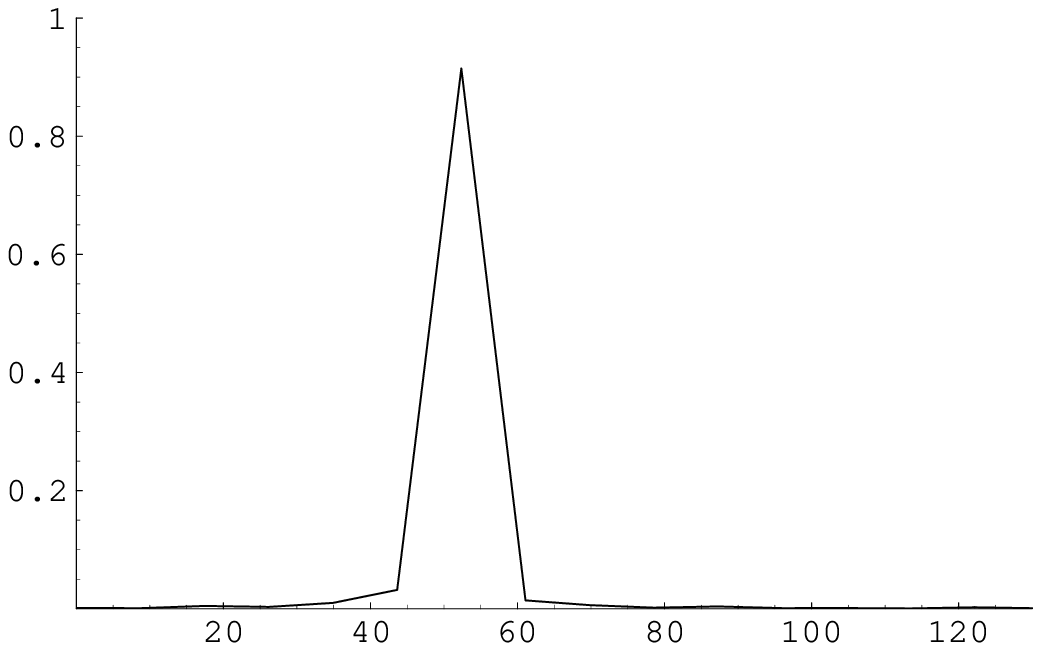}}&
\subfigure[]{\includegraphics[width=.4\textwidth]{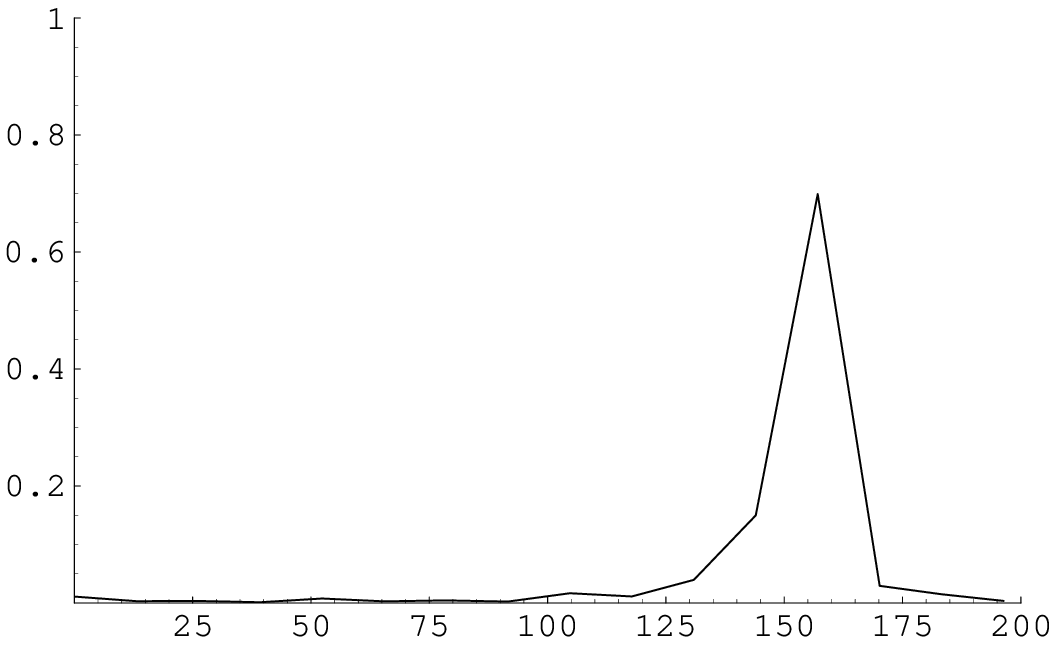}}
\end{tabular}
\end{center}
\caption{
Probability spectrum. (a) $\psi_0(x)=e^{-\omega x^2/2}$.
Parameters $t=0.045,\ n=30$.
(b) $\psi_0(x)=xe^{-\omega x^2/2}$.
Parameters $t=0.03,\ n=20$.
}
\end{figure*}

Fig.3 shows the result of the initial state
$\psi_0(x)=x^2e^{-\omega x^2/2}$.

\begin{figure}[h t b p]
 \begin{center}
  \includegraphics[width=.4\textwidth]{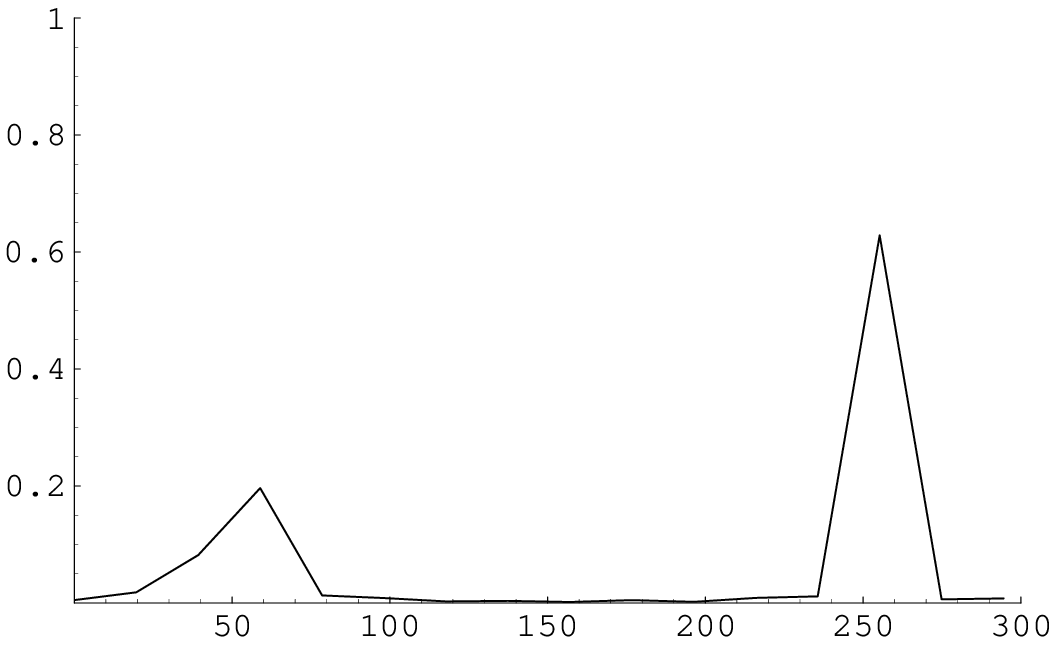}
 \end{center}
\caption{
Probability spectrum with $\psi_0(x)=x^2e^{-\omega x^2/2}$.
Parameters $t=0.02,\ n=20.$
}
\end{figure}

Although this is not the exact eigenstate,
it is a superposition of the ground state $\phi_0(x)$
and the second excited state $\phi_2(x)$,
i.e.,
\begin{equation}
\psi_0(x) \propto
\sqrt{\frac13}\phi_0(x)+\sqrt{\frac23}\phi_2(x)
\ .
\end{equation}
Therefore the energy spectrum shows two peaks
around $E \simeq 50,\ 250$, and the ratio
of the probability is roughly $1:2$, as is expected.

Figs.4 show the result of the initial state
$\psi_0(x)=1/\cosh^2(20x)$.
This state is a superposition of
even eigenstates. Fig.4(a) shows the peak at $E=52.4$
with probability $|c|^2=0.61$,
which corresponds to the ground state component.
The exact overlap value is $|c|^2=0.68$,
which is in good agreement.
The small bump around $E \simeq 110$ may come from higher
excited states.
Fig.4(b) shows the spectrum where the searched energy range is
extended up to $E=300$,
and the second peak at $E \simeq 250$ can be seen clearly .
The ratio of the probability is about $4:1$, which is also
in good agreement with the exact value.

\begin{figure*}[h t b p]
 \begin{center}
\begin{tabular}[t]{cc}
\subfigure[]{\includegraphics[width=.4\textwidth]{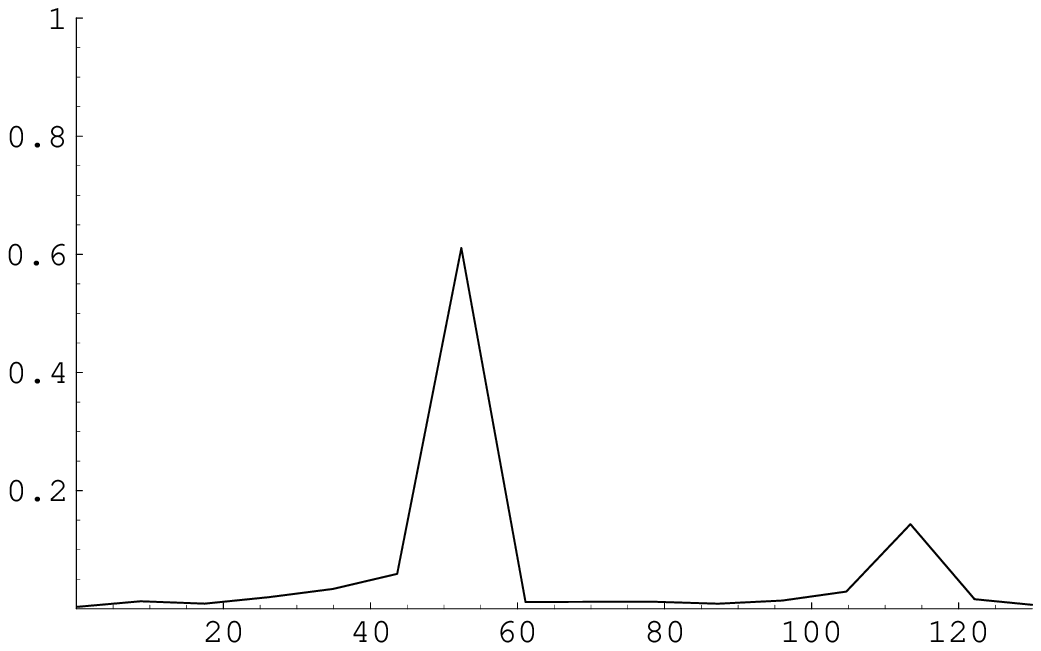}}&
\subfigure[]{\includegraphics[width=.4\textwidth]{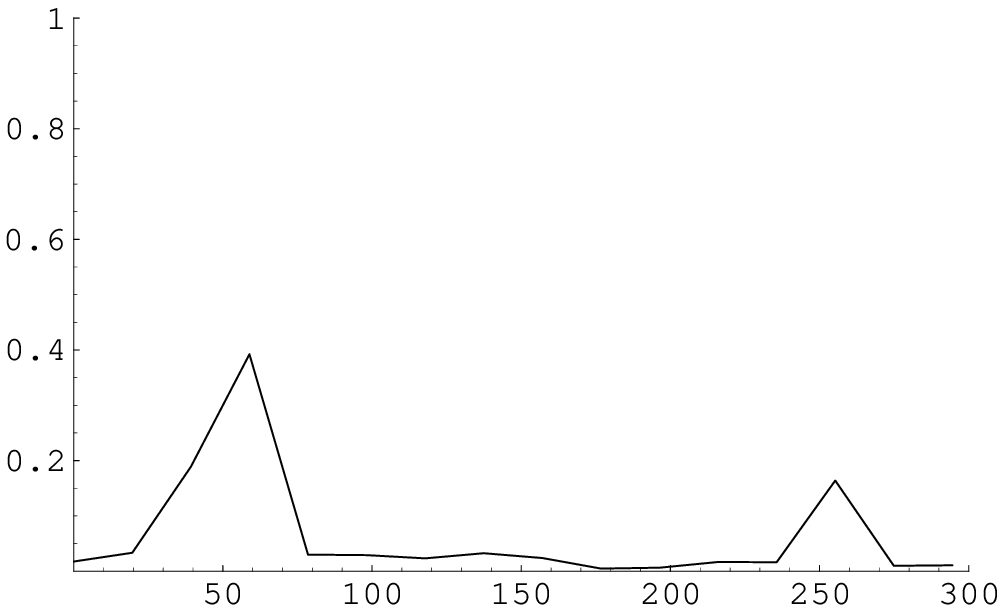}}
\end{tabular}
 \end{center}
\caption{
Probability spectrum with $\psi_0(x)=1/\cosh^2(20x)$.
(a) Parameters $t=0.045,\ n=30$.
(b) Parameters $t=0.02,\ n=20$.
}
\end{figure*}

Figs.5 show the projected eigenfunction corresponding
to the peak at energy $E=52.4$ of Fig.2(a).

\begin{figure*}[h t b p]
 \begin{center}
\begin{tabular}[t]{cc}
\subfigure[]{\includegraphics[width=.4\textwidth]{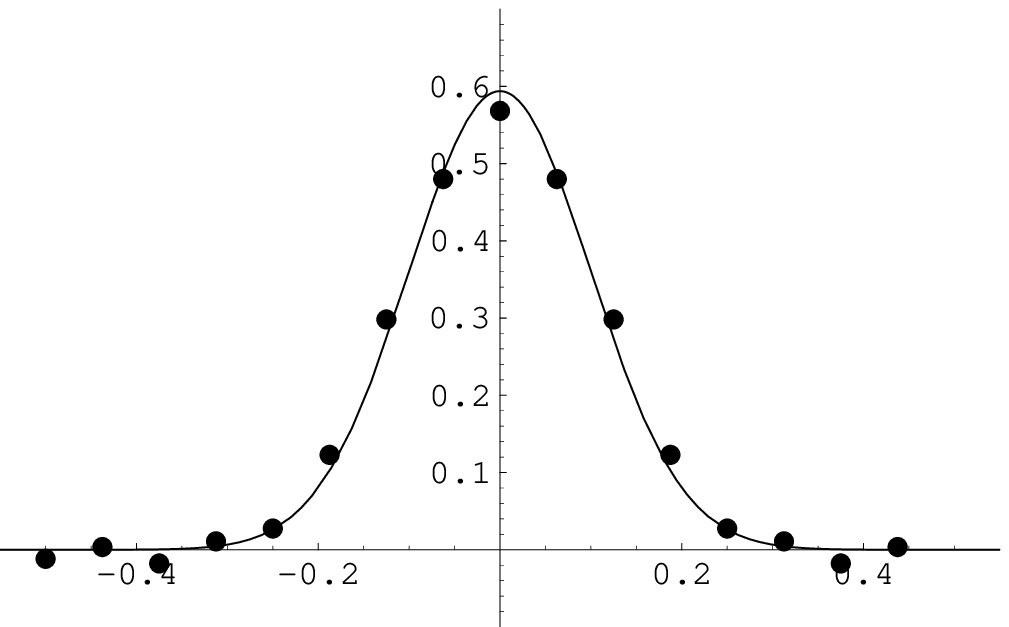}}&
\subfigure[]{\includegraphics[width=.4\textwidth]{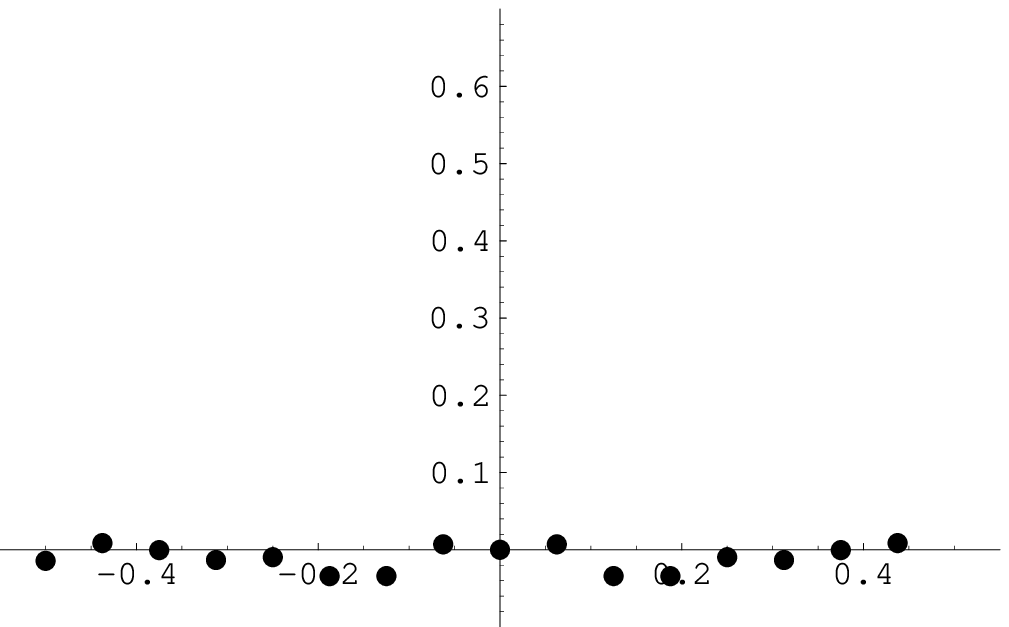}}
\end{tabular}
 \end{center}
\caption{
Projected eigenfunction of the ground state.
Solid line shows the exact wave function.
(a) Real part, (b) Imaginary part.
}
\end{figure*}

The wave function is normalized to be real at $x=0$. 
The solid line is the exact eigenfunction $\phi_0(x)$.
Since the initial state is an exact eigenstate,
the good agreement means that the QFT and the phase estimation
algorithm work properly.
The magnitude of imaginary part
shows the inaccuracy of this simulation.

One may wonder the outcome if the initial state is chosen randomly,
which might correspond to {\it ab initio} calculation.
Fig.6 shows the average of the outputs of 10 random initial states.
There are three broad peaks corresponding to the
exact energy values. This simulation shows that the initial state
should be prepared carefully.

\begin{figure}[h t b p]
 \begin{center}
  \includegraphics[width=.4\textwidth]{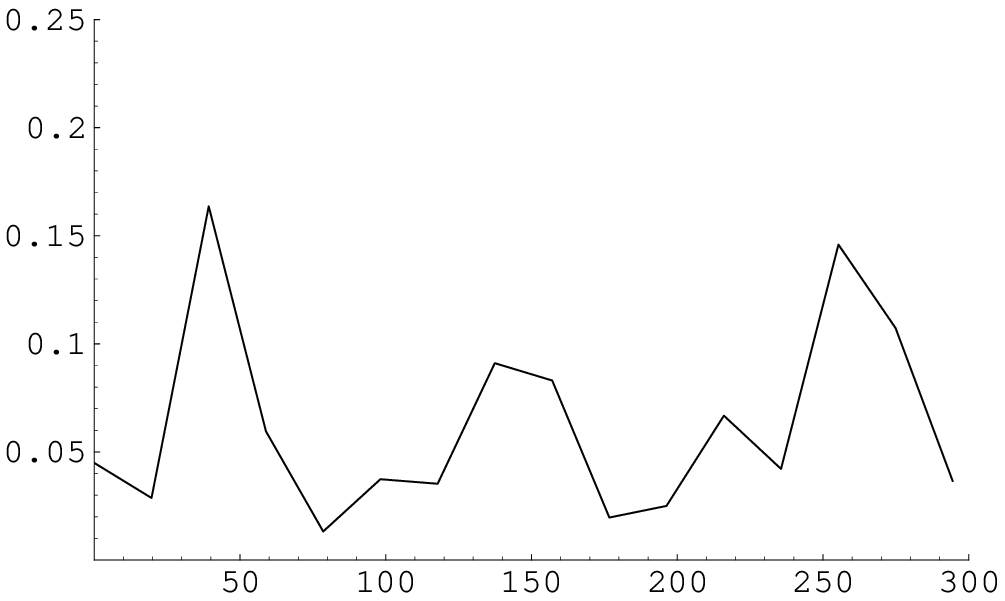}
 \end{center}
\caption{
Average of 10 random initial states.
}
\end{figure}

The example of the harmonic oscillator shows that the quadratic
potential can be constructed by single- and two-qubit operators.
One can readily understand that, for general $n$-th order
polynomial potential, the quantum circuits are given by
2-,\ 3-,\ldots,$n$-qubit operators,
 i.e., $CU,CCU,\ldots,C^{n-1} U$ gates.

\subsection{Square-well potential}

The Hamiltonian of the square-well potential is
\begin{equation}
H=\frac12 p^2+V(x) \ ,
\end{equation}
where the potential energy $V(x)$ is given by
\begin{equation}
V(x) =
\left \{
\begin{array}{r l}
-V_0 & \qquad |x| < a \\
0 & \qquad |x| > a \ .
\end{array} 
\right.
\end{equation}
We will fix the potential strength $V_0=100$ and
the range $a=1/4$ hereafter. Since the mesh points are
distributed in $[-1/2,1/2]$, we choose these parameters such that
the wave function is localized in this region.
And it also makes the quantum circuit very simple, although
the modification for general case is straightforward. 

Using the binary representation of $k=j_1j_2 \ldots j_s$,
the coordinate $x_k$ is written by
\begin{equation}
x_k = k/N_s-1/2 = \sum_{n=1}^s j_n 2^{-n} -1/2 \ .
\end{equation}
Thus, the first two qubits determine the position of $x$, i.e.,
\begin{subequations}
\begin{eqnarray}
-1/2 \le x < -1/4 &{\rm for}&j_1 = 0 ,\  j_2 = 0 \ ,\\
-1/4 \le x < 0    &{\rm for}&j_1 = 0 ,\  j_2 = 1 \ ,\\
   0 \le x < 1/4  &{\rm for}&j_1 = 1 ,\  j_2 = 0 \ ,\\
 1/4 \le x < 1/2  &{\rm for}&j_1 = 1 ,\  j_2 = 1 \ .
\end{eqnarray}
\end{subequations}
Therefore the potential energy becomes
\begin{equation}
V=
\left \{
\begin{array}{lll}
0 & \quad j_1=j_2=0 &  {\rm or} \quad j_1=j_2=1 \\
-V_0 & \quad j_1=0,j_2=1 &   {\rm or} \quad j_1=1,j_2=0 \ .
\end{array} 
\right.
\end{equation}
The time-evolution of the potential term
$e^{-iV \Delta t/2}$
can be expressed by the two-qubit operator working only on the first
two qubits $|j_1j_2\rangle$ as
\begin{equation}
B = \left(
\begin{array}{c c c c}
1 & 0 & 0 & 0 \\
0 & e^{i V_0 \Delta t/2} & 0 & 0 \\
0 & 0 & e^{i V_0 \Delta t/2} & 0 \\
0 & 0 & 0 & 1
\end{array}
\right)
\ .
\end{equation}
This circuit $B$ is constructed with a single-qubit operator
\begin{equation}
\tilde{B} = \left(
\begin{array}{cc}
e^{i V_0 \Delta t/2} & 0 \\
0 & e^{i V_0 \Delta t/2}
\end{array}
\right)
\ ,
\end{equation}
and $X$-operator (NOT-circuit)
which exchanges the coefficients of a single-qubit as
\begin{equation}
X = 
\left(
\begin{array}{cc}
0 & 1 \\
1 & 0
\end{array}
\right)
\ ,
\end{equation}
and controlled-$U$ operator. Fig.7 shows the quantum circuit
executing two-qubit operator $B$,
where the empty circle indicates that the operation is applied
on the target qubit when the control qubit is set to $|0 \rangle$.
The symbol $\oplus$ shows the $X$-operator (NOT-circuit).
\\

\begin{figure*}
\begin{center}
\unitlength 0.1in
\begin{picture}( 54.1000,  9.0000)(  5.9000,-15.0000)
%
\special{pn 8}%
\special{ar 1000 800 50 50  0.0000000 6.2831853}%
%
\special{pn 8}%
\special{pa 1200 600}%
\special{pa 1600 600}%
\special{pa 1600 1000}%
\special{pa 1200 1000}%
\special{pa 1200 600}%
\special{fp}%
%
\special{pn 8}%
\special{ar 1800 800 50 50  0.0000000 6.2831853}%
%
\special{pn 8}%
\special{ar 1400 1400 50 50  0.0000000 6.2831853}%
%
\special{pn 8}%
\special{pa 600 800}%
\special{pa 950 800}%
\special{fp}%
\special{pa 1050 800}%
\special{pa 1200 800}%
\special{fp}%
\special{pa 1600 800}%
\special{pa 1750 800}%
\special{fp}%
\special{pa 1850 800}%
\special{pa 2200 800}%
\special{fp}%
%
\special{pn 8}%
\special{pa 600 1400}%
\special{pa 1350 1400}%
\special{fp}%
\special{pa 1450 1400}%
\special{pa 2200 1400}%
\special{fp}%
%
\special{pn 8}%
\special{ar 1000 1400 50 50  0.0000000 6.2831853}%
%
\special{pn 8}%
\special{ar 1800 1400 50 50  0.0000000 6.2831853}%
%
\special{pn 8}%
\special{pa 1400 1000}%
\special{pa 1400 1350}%
\special{fp}%
%
\special{pn 8}%
\special{pa 2400 1070}%
\special{pa 2600 1070}%
\special{fp}%
\special{pa 2400 1130}%
\special{pa 2600 1130}%
\special{fp}%
%
\special{pn 8}%
\special{pa 3000 700}%
\special{pa 3200 700}%
\special{pa 3200 900}%
\special{pa 3000 900}%
\special{pa 3000 700}%
\special{fp}%
%
\special{pn 8}%
\special{sh 0.600}%
\special{ar 3400 800 50 50  0.0000000 6.2831853}%
%
\special{pn 8}%
\special{ar 3400 1400 50 50  0.0000000 6.2831853}%
%
\special{pn 8}%
\special{pa 3600 700}%
\special{pa 3800 700}%
\special{pa 3800 900}%
\special{pa 3600 900}%
\special{pa 3600 700}%
\special{fp}%
%
\special{pn 8}%
\special{pa 4000 1300}%
\special{pa 4200 1300}%
\special{pa 4200 1500}%
\special{pa 4000 1500}%
\special{pa 4000 1300}%
\special{fp}%
%
\special{pn 8}%
\special{sh 0.600}%
\special{ar 4400 1400 50 50  0.0000000 6.2831853}%
%
\special{pn 8}%
\special{pa 4600 1300}%
\special{pa 4800 1300}%
\special{pa 4800 1500}%
\special{pa 4600 1500}%
\special{pa 4600 1300}%
\special{fp}%
%
\special{pn 8}%
\special{pa 4200 600}%
\special{pa 4600 600}%
\special{pa 4600 1000}%
\special{pa 4200 1000}%
\special{pa 4200 600}%
\special{fp}%
%
\special{pn 8}%
\special{pa 5000 700}%
\special{pa 5200 700}%
\special{pa 5200 900}%
\special{pa 5000 900}%
\special{pa 5000 700}%
\special{fp}%
%
\special{pn 8}%
\special{pa 5600 700}%
\special{pa 5800 700}%
\special{pa 5800 900}%
\special{pa 5600 900}%
\special{pa 5600 700}%
\special{fp}%
%
\special{pn 8}%
\special{ar 5400 1400 50 50  0.0000000 6.2831853}%
%
\special{pn 8}%
\special{sh 0.600}%
\special{ar 5400 800 50 50  0.0000000 6.2831853}%
%
\special{pn 8}%
\special{pa 2800 800}%
\special{pa 3000 800}%
\special{fp}%
\special{pa 3200 800}%
\special{pa 3350 800}%
\special{fp}%
\special{pa 3450 800}%
\special{pa 3600 800}%
\special{fp}%
\special{pa 3800 800}%
\special{pa 4200 800}%
\special{fp}%
%
\special{pn 8}%
\special{pa 4600 800}%
\special{pa 5000 800}%
\special{fp}%
\special{pa 5200 800}%
\special{pa 5350 800}%
\special{fp}%
\special{pa 5450 800}%
\special{pa 5600 800}%
\special{fp}%
\special{pa 5800 800}%
\special{pa 6000 800}%
\special{fp}%
%
\special{pn 8}%
\special{pa 2800 1400}%
\special{pa 4000 1400}%
\special{fp}%
\special{pa 4200 1400}%
\special{pa 4350 1400}%
\special{fp}%
\special{pa 4450 1400}%
\special{pa 4600 1400}%
\special{fp}%
\special{pa 4800 1400}%
\special{pa 6000 1400}%
\special{fp}%
%
\special{pn 8}%
\special{pa 3400 850}%
\special{pa 3400 1450}%
\special{fp}%
%
\special{pn 8}%
\special{pa 4400 1000}%
\special{pa 4400 1350}%
\special{fp}%
%
\special{pn 8}%
\special{pa 5400 850}%
\special{pa 5400 1450}%
\special{fp}%
%
\special{pn 8}%
\special{pa 1000 850}%
\special{pa 1000 1450}%
\special{fp}%
%
\special{pn 8}%
\special{pa 1800 850}%
\special{pa 1800 1450}%
\special{fp}%
\put(14.0000,-8.0000){\makebox(0,0){\LARGE $\tilde{B}$}}%
\put(44.0000,-8.0000){\makebox(0,0){\LARGE $\tilde{B}$}}%
\put(31.0000,-8.0000){\makebox(0,0){X}}%
\put(41.0000,-14.0000){\makebox(0,0){X}}%
\put(47.0000,-14.0000){\makebox(0,0){X}}%
\put(37.0000,-8.0000){\makebox(0,0){X}}%
\put(51.0000,-8.0000){\makebox(0,0){X}}%
\put(57.0000,-8.0000){\makebox(0,0){X}}%
\end{picture}%
\\

FIG.\ 7:\ Quantum circuit $B$
\end{center}
\end{figure*}
\setcounter{figure}{7}

Figs.8 show the probability
spectrum $|c_k|^2$ as a function of the energy $E$
for initial states $\psi_0(x)=e^{-10x^2}$ (even state) and
$\psi_0(x)=xe^{-10x^2}$ (odd state) respectively.

\begin{figure*}[h t b p]
 \begin{center}
\begin{tabular}[t]{cc}
\subfigure[]{\includegraphics[width=.4\textwidth]{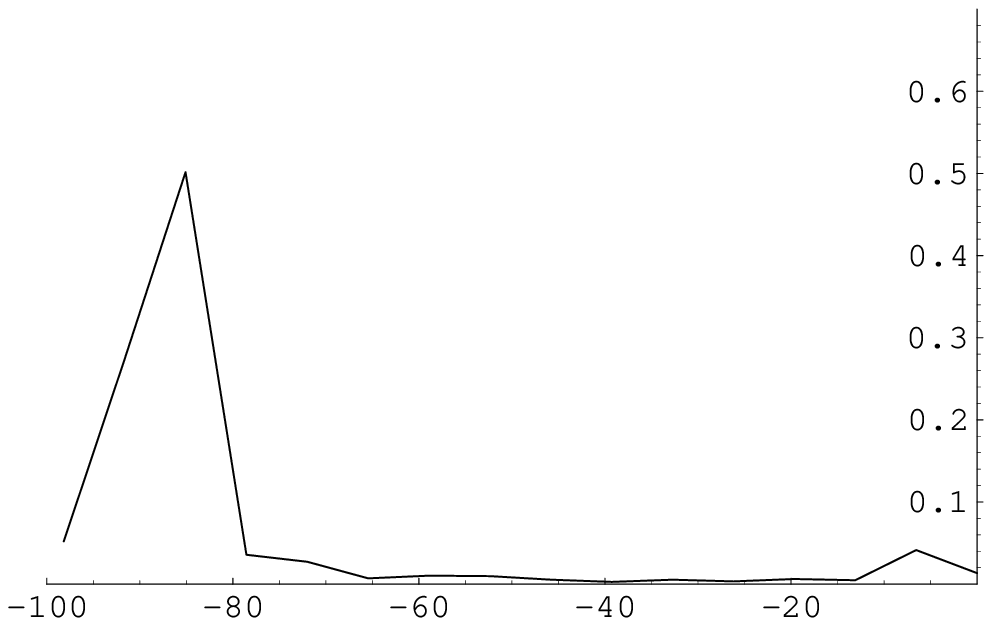}}&
\subfigure[]{\includegraphics[width=.4\textwidth]{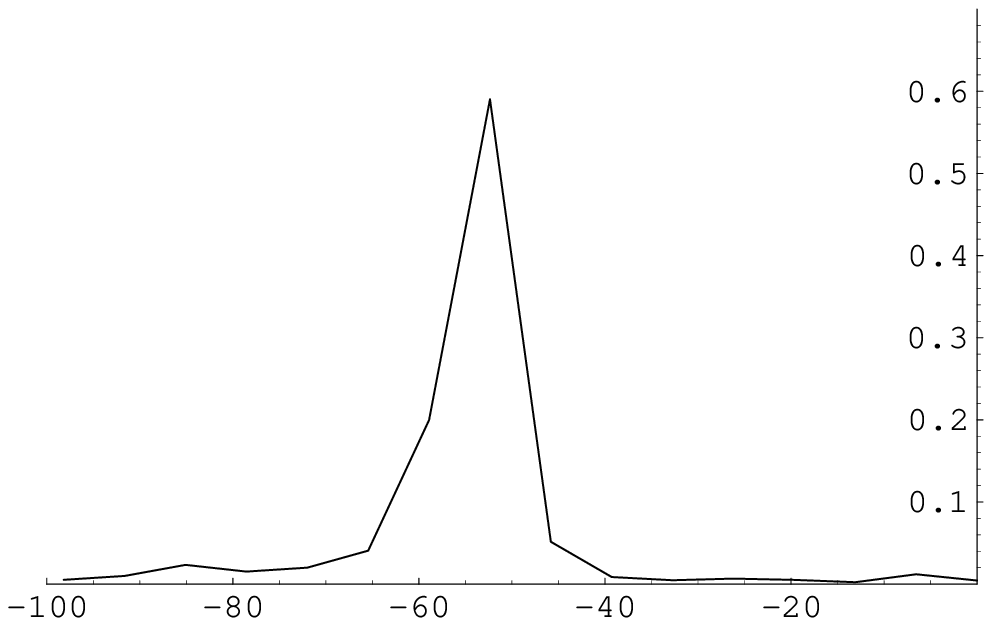}}
\end{tabular}
 \end{center}
\caption{
Probability spectrum. (a) $\psi_0(x)=e^{-10x^2}$.
Parameters $t=0.06,\ n=50$.
(b) $\psi_0(x)=xe^{-10x^2}$.
Parameters $t=0.06,\ n=50$.
}
\end{figure*}

The exact energy levels are $E_0=-88.12$,
$E_1=-54.05$ and $E_2=-7.005$.
Fig.8(a) shows a sharp peak at $E \simeq -85$ corresponding
to the ground state,
while a small bump at $E \simeq -7$ corresponds to the second excited state.
Fig.8(b) also shows a sharp peak at $E \simeq -55$ corresponding to the first
excited state.

Figs.9 shows the projected wave function corresponding to the energy
$E=-85.08$ of Fig.8(a).

\begin{figure*}[h t b p]
 \begin{center}
\begin{tabular}[t]{cc}
\subfigure[]{\includegraphics[width=.4\textwidth]{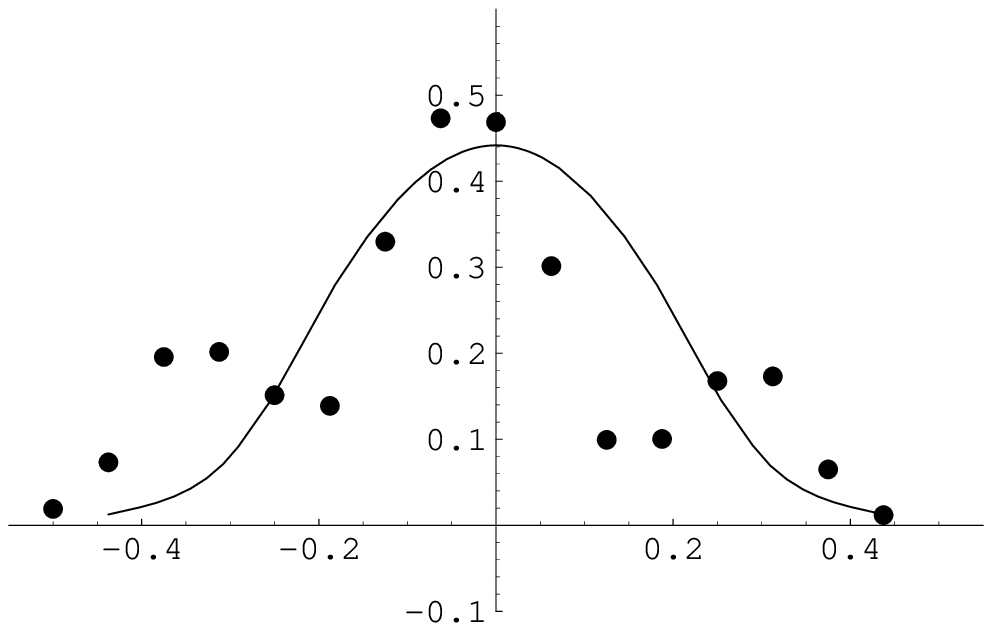}}&
\subfigure[]{\includegraphics[width=.4\textwidth]{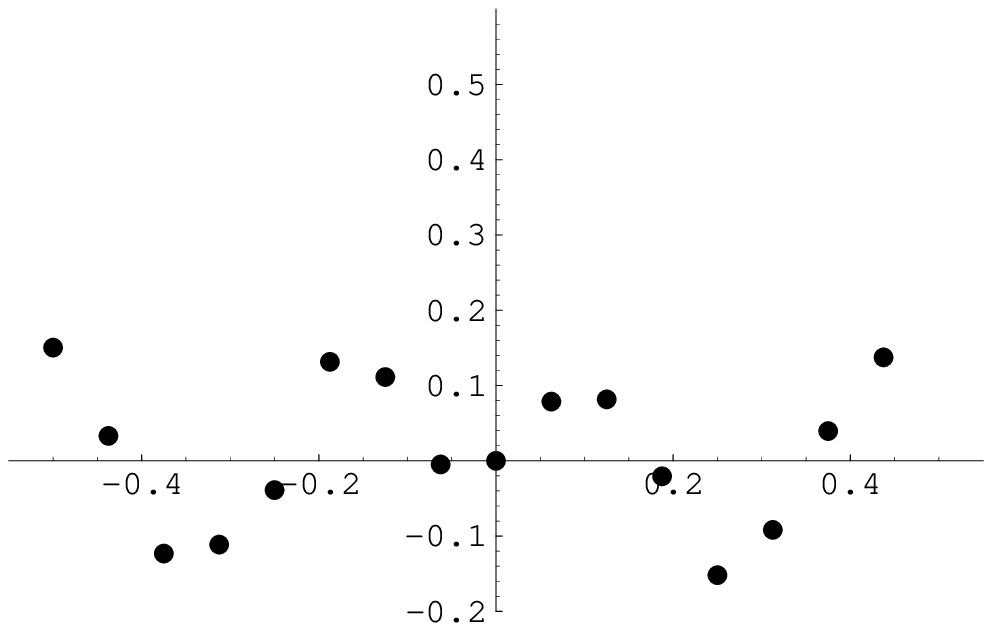}}
\end{tabular}
 \end{center}
\caption{
Projected eigenfunction of the ground state.
Solid line shows the exact wave function.
(a) Real part, (b) Imaginary part.
}
\end{figure*}

The phase of the wave function is normalized as
${\rm Im}(\psi(0))=0$. The exact wave function is shown by a solid line.
The agreement is not so good as compared with the harmonic oscillator
case. The wave function is slightly asymmetric,\ i.e., shifted
to the negative direction, and also shows a strange behavior
at $|x| \simeq 0.3$. The mixture of the imaginary part is not
small, which clearly shows that the simulation has some problems.
This is mainly caused by the fact that the potential is not exactly
symmetric. This is due to the asymmetric distribution of the
mesh points $\{x_k\}$. Namely, at the boundaries of the potential-well,
the strength is $V=-V_0$ at $x=-1/4$ corresponding
to $|j_1j_2\rangle=|01\rangle$, while $V=0$ at $x=1/4$
corresponding to $|j_1j_2\rangle=|11\rangle$. Therefore the potential-well
is negatively shifted by $\delta x=1/2^5$ in this case.
Another reason may be due to the sharp change of the potential
at the boundary.

In order to improve the simulation,
we have employed the symmetric distribution of the mesh points
given by Eq.(\ref{eq14}), which also makes the potential exactly symmetric.
The result is shown in Fig.10 and Figs.11.

\begin{figure}[h t b p]
 \begin{center}
  \includegraphics[width=.4\textwidth]{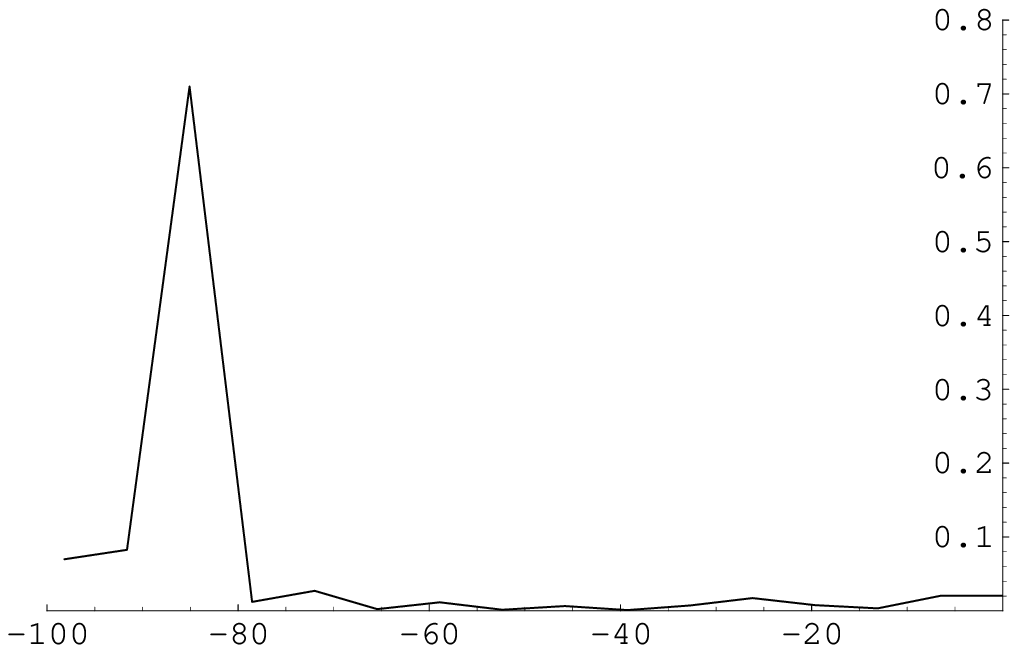}
 \end{center}
\caption{
Probability spectrum with $\psi_0(x)=e^{-10x^2}$.
Parameters $t=0.06,\ n=50$.
Mesh points are symmetrically distributed.
}
\end{figure}

\begin{figure*}[h t b p]
 \begin{center}
\begin{tabular}[t]{cc}
\subfigure[]{\includegraphics[width=.4\textwidth]{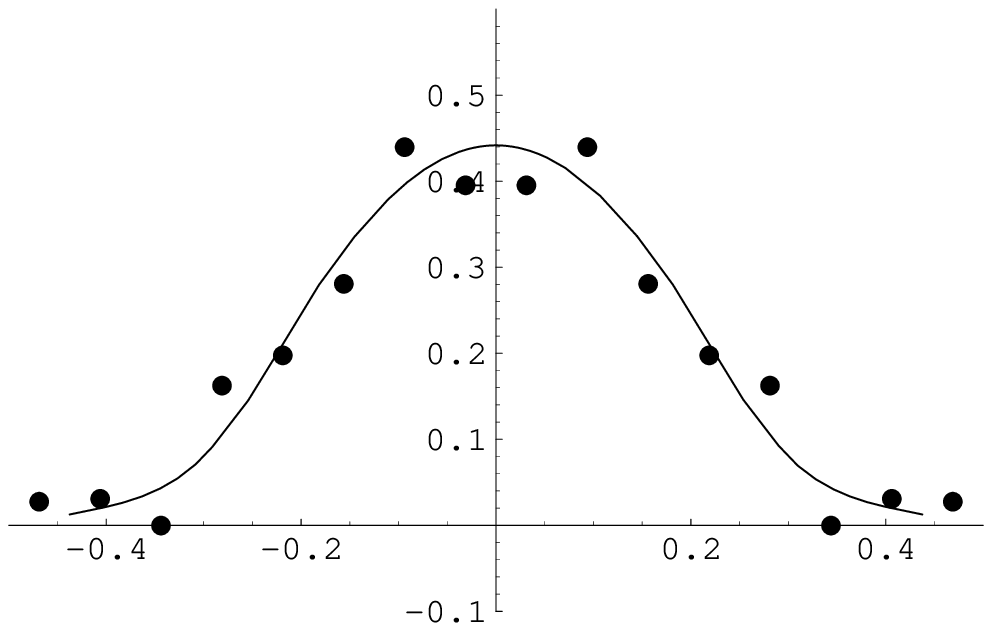}}&
\subfigure[]{\includegraphics[width=.4\textwidth]{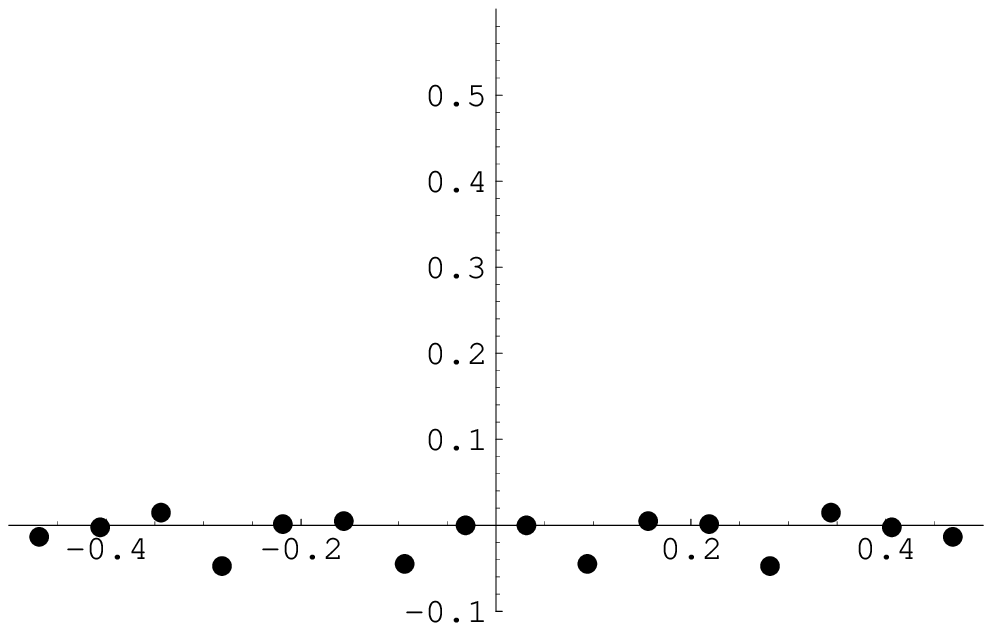}}
\end{tabular}
 \end{center}
\caption{
Projected eigenfunction of the ground state
with symmetric mesh points.
Solid line shows the exact wave function.
(a) Real part, (b) Imaginary part.
}
\end{figure*}

Now the probability shows
the more pronounced peak at $E=-85.08$.
The phase of the wave function is set to real
at $|x|=1/2^5$.
Figs.11 show that the agreement of the calculated wave function
with the exact values becomes much better.

Fig.12 shows the average result of 10 random initial states.
Although two lowest states ($E \simeq -88,\ -54$) may be seen
as broad peaks, the third state ($E \simeq -7$) cannot be resolved,
and small fractions of many eigenstates seem fill
over wide energy range.

\begin{figure}[h t b p]
 \begin{center}
  \includegraphics[width=.4\textwidth]{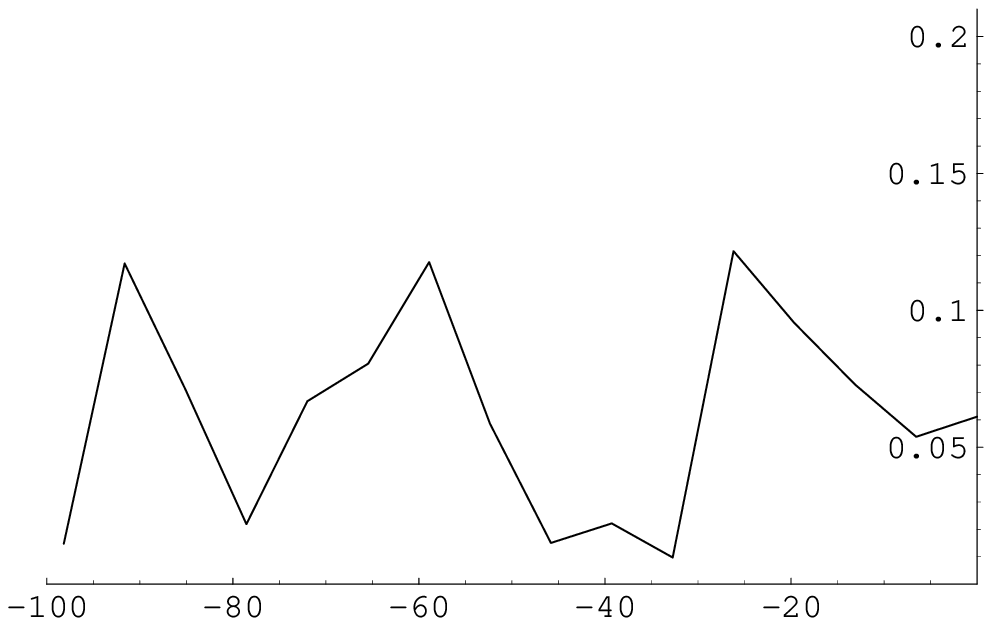}
 \end{center}
\caption{
Average of 10 random initial states.
}
\end{figure}

\subsection{Coulomb potential}

The Hamiltonian of the Coulomb potential is
\begin{equation}
H=\frac12 p^2 - \frac{\kappa}{r} \ ,\quad  (\kappa > 0) \ .
\end{equation}
The Schr\"odinger equation
is reduced to one-dimensional equation in the case of $S$-wave.
Thus the solution $\psi(r)$ is given
by $\psi(r)=rR_0(r)\ (r \ge 0)$,
where $R_0(r)$ is the radial part of the $S$-wave
Coulomb wave function. If the potential $V(x)=-\kappa/|x|$
is defined in $-\infty < x < \infty$, the energy eigenvalues
are doubly degenerate except for the ground state.
We will take the odd wave function $\psi(x)=xR_0(|x|)$
by setting odd initial states, since it is smooth at $x=0$.
The problems of the one-dimensional Coulomb
potential have been discussed
in Refs.\cite{lou,imb} in detail.

The construction of the quantum circuit of the Coulomb
potential is not straightforward,
since it is necessary to express
the inverse of the binary fraction.
We have made a simple expression in the following way.

Let $ 0 < x < 1$ be expressed as the binary fraction as
\begin{equation}
x= \sum_{k=1}^N j_k 2^{-k} = 0.j_1j_2 \ldots j_N \ ({\rm binary}) \ .
\end{equation}
We will find the formula $y=1/x$ in terms
of $j_1,j_2,\ldots,j_N$. The first bit $j_1$ can be set
$j_1=1\ (1/2 \le x < 1)$.
In the case of
$j_1=0\ (0 < x < 1/2)$, one can shift the binary expression
by an appropriate power of 2. Since $y=1/x > 1$,
$y$ can be expressed by the power series of $1/2$ as
\begin{equation}
y=1+\sum_{\ell\geq 1} a_\ell 2^{-\ell} \ .
\label{eq44}
\end{equation}
Note that the integer coefficients $a_\ell$ are not
necessarily 0 or 1. In fact we find that
$a_\ell{}'s$ are small integers and the Eq.(\ref{eq44})
converges rapidly.

The coefficients $a_\ell$ are determined by the equation
\begin{eqnarray}
x \times y
&=& \Bigl(\sum_{k=1}^N j_k 2^{-k}\Bigr)
\Bigl(1+\sum_{\ell\geq 1} a_\ell 2^{-\ell}\Bigr)
\nonumber \\
&=& \sum_{k=1}^N j_k 2^{-k} +\sum_{k,\ell}j_k a_\ell 2^{-k-\ell}
\nonumber \\
&=& 1 \ .
\end{eqnarray}
Since $1=1/2 + 1/2^2 + 1/2^3 + \ldots=0.111\ldots$
in the binary fraction, one can obtain the equations
which determine the coefficients $a_\ell$ recursively,
\begin{equation}
j_m + \sum_{k+\ell=m} j_ka_\ell = 1 , \quad m=2,3,\ldots \ .
\end{equation}
Up to $N=7$, $a_\ell{}'s$ are expressed as follows,
\begin{subequations}
\label{eq47}
\begin{eqnarray}
a_1 &=& 1-j_2 \ ,\\
a_2 &=& 1-j_3 \ ,\\
a_3 &=& 1-j_2-j_3-j_4+2j_2j_3 \ ,\\
a_4 &=& 1-j_4-j_5-j_2j_3+2j_2j_4 \ ,\\
a_5 &=& 1-j_2-j_4-j_5-j_6-j_2j_4+2j_2j_5+2j_3j_4 \ ,\\
a_6 &=& 1-j_3-j_5-j_6-j_7 +j_2j_3+3j_2j_4-j_2j_5 \nonumber \\
    & & +2j_2j_6+2j_3j_4+2j_3j_5-6j_2j_3j_4 \ .
\end{eqnarray}
\end{subequations}
In the case $j_1=0$, one can obtain similar expressions
by shifting $j_m \to j_{m+1}$ and multiplying by 2.

The Coulomb potential is an even function
and it has a singular point $x=0$.
Therefore the exactly symmetric mesh points of Eq.(\ref{eq14})
is suitable.
For the case of simulation qubits $s=4$, mesh points are
explicitly given by
\begin{equation}
x_k=x-(1/2-1/2^5)=0.j_1j_2j_3j_4-0.1+0.00001 \ ({\rm binary})
\ .
\end{equation}
The potential is proportional to the inverse of
the absolute value $|x_k|$, which is given by
\begin{equation}
|x_k| =
\left \{
\begin{array}{ll}
0.0j_2j_3j_41\ ({\rm binary}) & \quad {\rm for}\ j_1=1 \\
0.0j'_2j'_3j'_41\ ({\rm binary}) & \quad {\rm for}\ j_1=0 \ ,
\end{array} 
\right.
\end{equation}
where $j'_m=1-j_m$ is the bit-flip of $j_m$.
Note that we can formally set $j_5=1$ for both cases.
Thus, for $x_k<0$, one should apply bit-flip operation
before executing the time-evolution operator.

The time-evolution operator of the potential term
$e^{-iV\Delta t/2}$ can be
constructed recursively depending on
whether the qubit is $|0 \rangle$ or $|1 \rangle$.
Defining the projection operator $P_0\ (P_1)$ to the qubit
$|0 \rangle\ (|1 \rangle)$,
\begin{equation}
\begin{array}{cc}
P_0 = 
\left(
\begin{array}{cc}
1 & 0 \\
0 & 0
\end{array}
\right), \ 
&
P_1 = 
\left(
\begin{array}{cc}
0 & 0 \\
0 & 1
\end{array}
\right)
\\
\end{array}
\ ,
\end{equation}
the matrix $U_1\ (2^3\times 2^3)$ corresponding
to $x_k>0$ is given by
\begin{subequations}
\begin{eqnarray}
U_1 &=& P_0(j_2) \otimes U_2
 + P_1(j_2) \otimes e^{i\kappa V_2(j_3,j_4)\Delta t/2 } \ ,\\
U_2 &=& P_0(j_3) \otimes U_3
 + P_1(j_3) \otimes e^{i\kappa V_3(j_4)\Delta t/2} \ ,\\
U_3 &=& P_0(j_4) \otimes e^{i\kappa U_4\Delta t/2}
 + P_1(j_4) \otimes e^{i\kappa V_4\Delta t/2} \ ,
\end{eqnarray}
\end{subequations}
with
\begin{subequations}
\begin{eqnarray}
V_2(j_3,j_4)
&=& 
2\{1+(1-j_3)2^{-1}+(1-j_4)2^{-2}+(-j_3-j_4+2j_3j_4)2^{-3}
 \nonumber \\
& & \ +(2j_3-j_3j_4)2^{-4}+(-2j_3+2j_4)2^{-5}
+(1+3j_3+j_4-5j_3j_4)2^{-6}
\}
\ ,\\
V_3(j_4)
&=&
2^2\{1+(1-j_4)2^{-1}+j_42^{-3}+(1-j_4)2^{-4}
+(1-j_4)2^{-5}+j_42^{-6}
\}
\ ,\\
V_4
&=&
2^3(1+2^{-2}+2^{-4}+2^{-6} )
\ ,\\
U_4
&=&
2^4(1+2^{-1}+2^{-2}+2^{-3}+2^{-4}+2^{-5}+2^{-6} ) \ .
\end{eqnarray}
\end{subequations}
These formulas can be obtained by appropriately modifying
the basic formula Eqs.(\ref{eq47}).
The time-evolution operators are single- or two-qubit operators,
and can be constructed in the same way
as the harmonic oscillator case.

The simulations are carried out with a strength parameter
$\kappa=10$.
The accuracy of our approximation of the Coulomb potential
with $s=4$ simulation qubits is within 1.6\%,
which might be sufficient for simulations.

Fig.13 shows the probability spectrum as a function
of the energy $E$ for the exact initial state
$\psi_0(x)=xe^{-10|x|}$.

\begin{figure}[h t b p]
 \begin{center}
  \includegraphics[width=.4\textwidth]{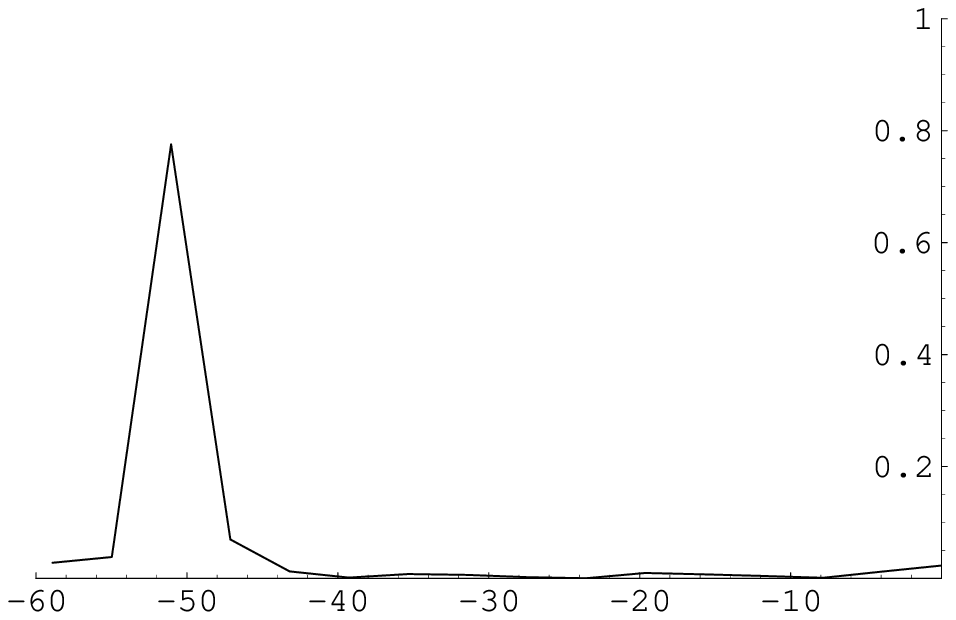}
 \end{center}
\caption{
Probability spectrum with $\psi_0(x)=xe^{-10|x|}$.
Parameters $t=0.1,\ n=100$.
}
\end{figure}

The exact energy is $E_0=-\kappa^2/2=-50$, and
the agreement is satisfactory.
Fig.14 shows the energy spectrum with initial state
$\psi_0(x)=x|x|e^{-10|x|}$, which contains excited states.

\begin{figure}[h t b p]
 \begin{center}
  \includegraphics[width=.4\textwidth]{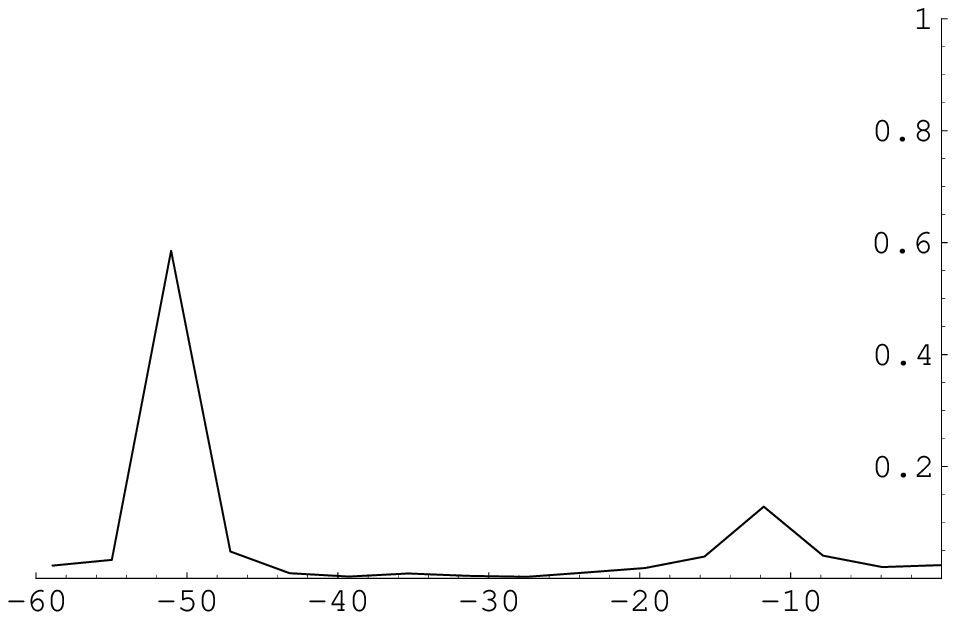}
 \end{center}
\caption{
Probability spectrum with $\psi_0(x)=x|x|e^{-10|x|}$.
Parameters $t=0.1,\ n=100$.
}
\end{figure}

The spectrum shows another bump around $E \simeq -10$, which
corresponds to the first excited state with energy
$E_1=-\kappa^2/8=-12.5$.
Fig.15 shows the projected wave function corresponding to
$E=-51.05$ of Fig.13.

\begin{figure*}[h t b p]
 \begin{center}
\begin{tabular}[t]{cc}
\subfigure[]{\includegraphics[width=.4\textwidth]{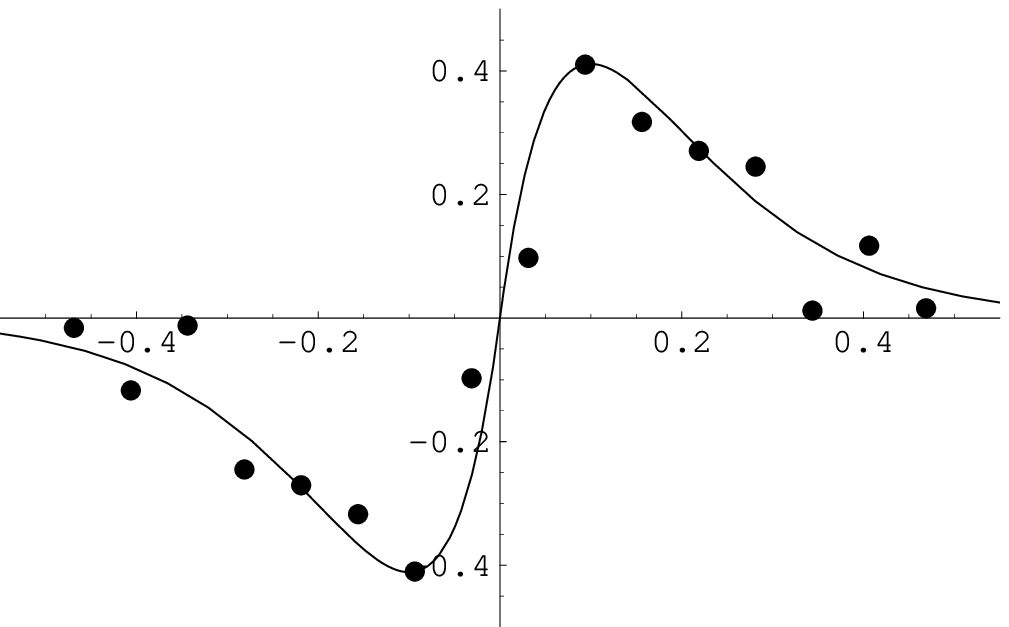}}&
\subfigure[]{\includegraphics[width=.4\textwidth]{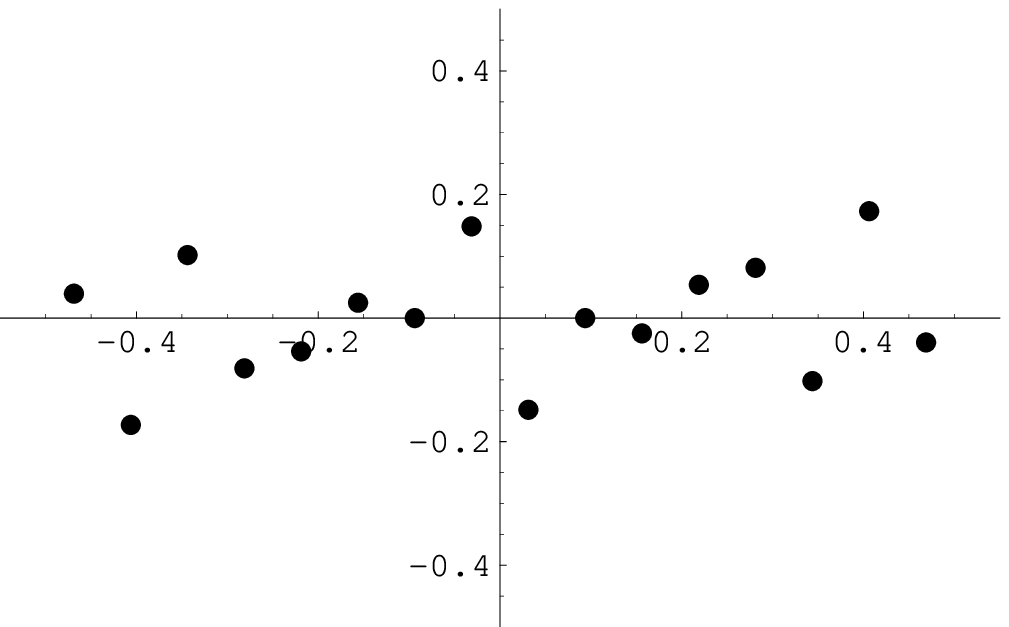}}
\end{tabular}
 \end{center}
\caption{
Projected eigenfunction of the ground state.
Solid line shows the exact wave function.
(a) Real part, (b) Imaginary part.
}
\end{figure*}

The phase of the wave function is set to real
at the maximum amplitude ($|x|=3/2^5$). The agreement
seems fairly good, although the mixture of the imaginary part
is not negligible.

Fig.16 shows the average result of 10 random initial states.
In this case, only the ground state ($E \simeq -50$) can be seen.
This is because excited states are accumulated near $E \simeq 0$
in the Coulomb potential, and positive energy continuum states
might contribute to fill the whole energy range
due to the periodicity.

\begin{figure}[h t b p]
 \begin{center}
  \includegraphics[width=.4\textwidth]{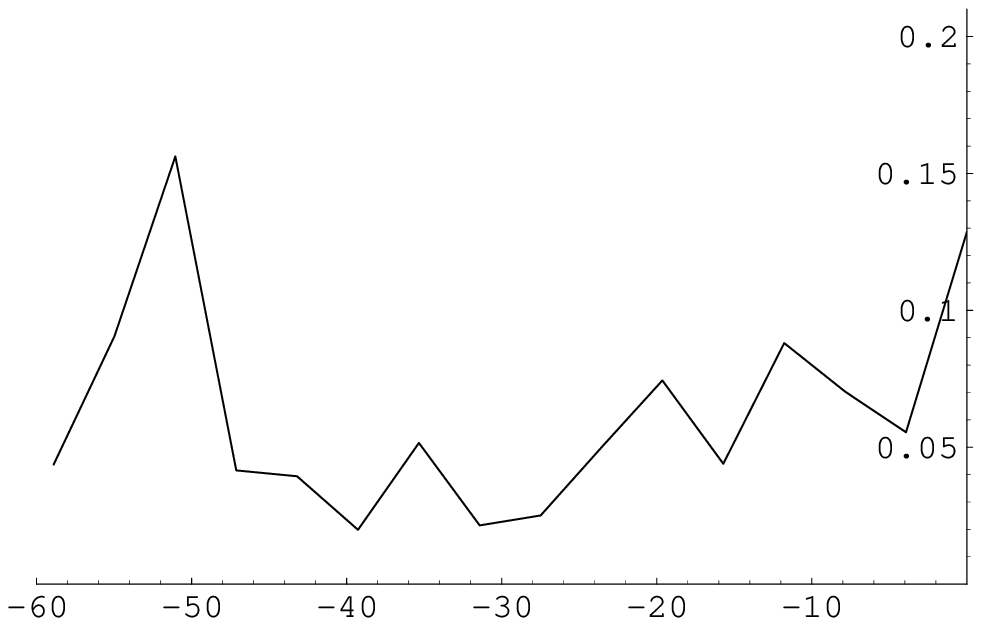}
 \end{center}
\caption{
Average of 10 random initial states.
}
\end{figure}

\section{Summary}

We have explicitly constructed quantum circuits and
carried out simulations of typical one-dimensional
Schr\"odinger equations, i.e., harmonic oscillator,
square-well and Coulomb potential.
We have made quantum circuits in such a way that
they consist of only single-qubit and two-qubit operators
and do not require ancillary qubits to calculate
the potential term. Therefore they are simple and easy
for implementation. With eight qubits (4 work qubits and
4 simulation qubits), our simulations could obtain reasonable
outputs compared with the exact results.
It is found that exactly symmetric mesh points should be
employed for the symmetric potential, and
the initial states should be prepared deliberately.

\end{document}